\documentclass[twocolumn,tightenlines,showpacs,aps]{revtex4}
\usepackage{graphicx}

\begin{document}

\title{Quantum Mechanical Hysteresis and the Electron Transfer Problem}

\author{P. G. Etchegoin}
\email{Pablo.Etchegoin@vuw.ac.nz}

\affiliation{The McDiarmid Institute for Advanced Materials and Nanotechnology\\
School of Chemical and Physical Sciences\\ Victoria University of Wellington\\
PO Box 600 Wellington, New Zealand}

\date{\today}

\begin{abstract}
We study a simple quantum mechanical symmetric donor-acceptor
model for electron transfer (ET) with coupling to internal
deformations. The model contains several basic properties found in
biological ET in enzymes and photosynthetic centers; it produces
tunnelling with hysteresis thus providing a simple explanation for
the slowness of the reversed rate and the near 100\% efficiency of
ET in many biological systems. The model also provides a
conceptual framework for the development of molecular electronics
memory elements based on electrostatic architectures.
\end{abstract}

\pacs{82.30.Fi; 82.39.Jn; 05.45.-a; Keywords: Tunnelling, Monte
Carlo, Electron transfer, Biological Processes}

\maketitle

The physics of electron transfer (ET) is of paramount importance
in many fundamental aspects of molecular biology, from
photosynthesis\cite{Hall}, to mitochondrial ET, to many control
reactions and signalling through the cell
membrane\cite{DeVault,Volkenstein,Davydov}. Being a sophisticated
quantum mechanical tunnelling process in itself, the ET problem
includes many aspects of coupling to vibrational degrees of
freedom, tunnelling control through external variables, and
non-reversibility aspects\cite{DeVault}. After the pioneering work
of Marcus\cite{Marcus} in the electrochemical aspects of ET for
redox reactions in solution, the problem has seen several times
renewed interest in both semiclassical and quantum mechanical (QM)
aspects. This is an area where there is a strong interplay between
QM and important biological effects at the molecular level.

In addition, there is considerable evidence now on the importance
of nonlinear contributions to many important dynamical aspects of
molecular biology. DNA dynamics and
denaturation\cite{Bishop1,Bishop2} and signal transduction and
coordination of events during the DNA readout by
proteins\cite{Yakusevich,Salerno} have both been ascribed to
nonlinearities. The nonlinear character of the hydrogen bonding
potential among DNA bases has been known for a very long
time\cite{Davydov,dna1,dna2}. But nonlinearities in biology are
believed to play also a substantial role in ET and energy
transfer. Resonant dipole-dipole interactions of the amide I band
combined with structural deformations in alpha-helical proteins
have been proposed for a very long time as a possible explanation
for energy transport from the hydrolysis of ATP\cite{Davi1}. There
is experimental evidence that amide I vibrations are coupled
nonlinearly; an example being their unusually long life in simple
proteins\cite{Xie}. The coupling of carriers to conformational
degrees of freedom has been suggested also to exist in connection
with ET from the active prosthetic groups inside
cytochromes\cite{Davydov}. There seems to be, in addition,
substantial evidence that a nonlinear coupling is required to
explain the so-called {\it gated ET reactions} between quinones in
the photosynthetic reaction centers\cite{Rabenstein}. Knapp and
coworkers\cite{Knapp} were among the first to realize the
importance of both nonlinearities and Davidov-like
solitons\cite{Davydov} in the ET problem. They argued that the
construction of the initial and final wavefunctions in terms of
solitary states leads automatically to localization for a
symmetric donor-acceptor system, a very desirable model property
in many cases, as we shall explain later.

The brief outline given above highlights the fact that there is a
case in many real biological systems for the study of QM
tunnelling properties in the presence of intrinsic nonlinearities.
To this end, there are two possible options, either we study the
specific non-linear coupling of a given case with all its
peculiarities and details, as done in Ref. \cite{Rabenstein} for
ET between quinones in the photosynthetic reaction center, or we
study a skeletal model which contains the essential elements of
the problem and try to draw general qualitative conclusions from
its physical behavior. We shall adopt the latter view here to
highlight a specific property of the tunnelling process in the
presence of nonlinearities; i.e. the presence of hysteresis in a
symmetric donor-acceptor ET system.

\begin{figure}[h]
 \centering{
  \includegraphics[width=8cm, height = 9cm]{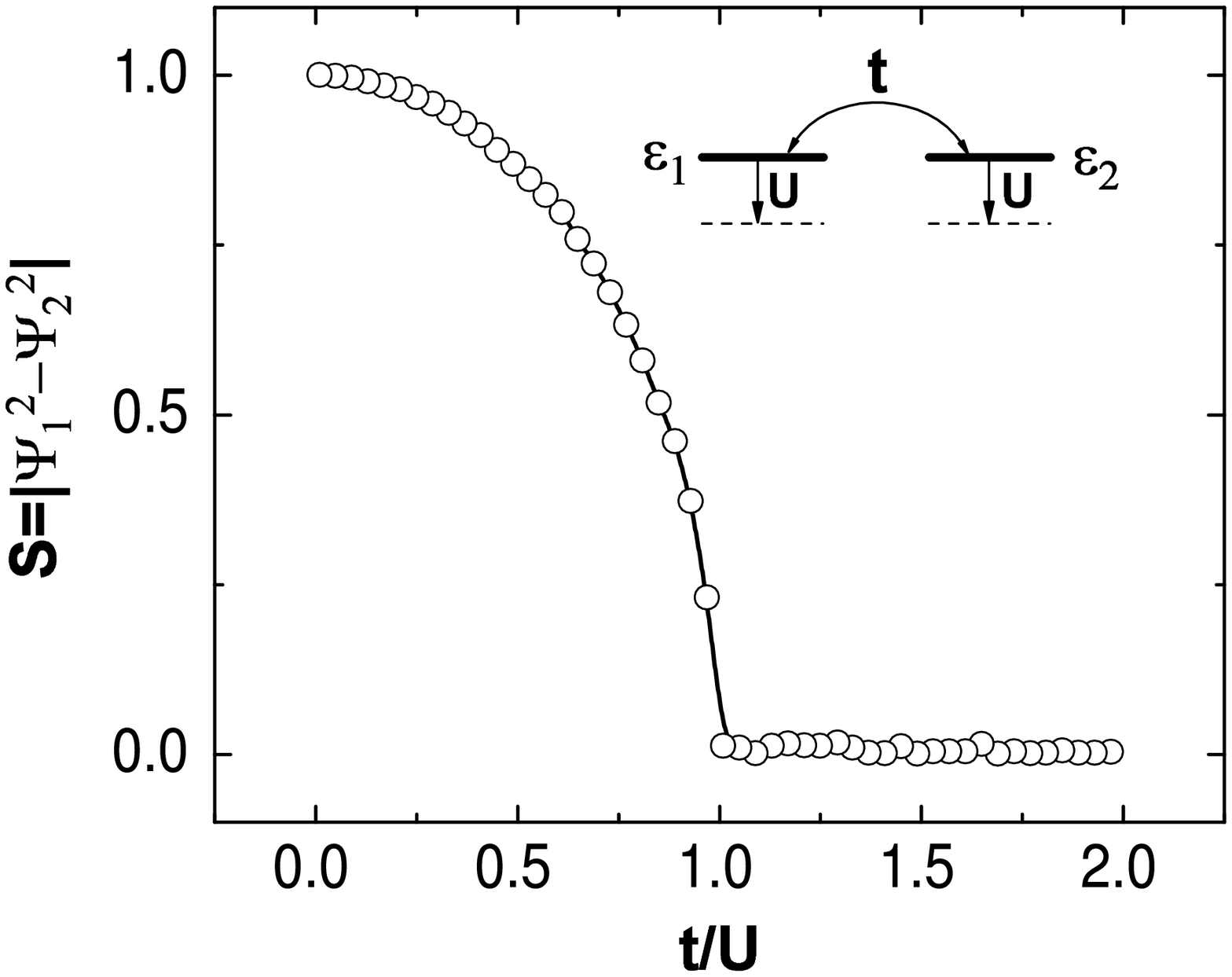}
 }
\caption{Asymmetry parameter $S$ (Eqn. \ref{eqn13}) for the ground
state as a function of $t/U$ for the two-site NLH model (Eqn.
\ref{eqn11}). A schematic view of the model and its parameters is
given in the inset. The calculation is for a symmetric
donor-acceptor pair with $\epsilon_1=\epsilon_2=0$. Below $t/U=1$
the system profits by accumulating the electron on one of the
sites, thus creating an asymmetric self-localized ground state.}
 \label{fig1}
\end{figure}

Davydov's approach to electron-lattice deformation coupling in
biological systems\cite{Davydov} and its variants\cite{Knapp} lead
always to an effective Hamiltonian of the non-linear Schr\"odinger
equation (NLS) type. The nonlinearity in the equation comes from a
{\it hidden} adiabatic coupling to an external interaction (the
deformation) which is factored out from the dynamics in the form
of an effective self-consistent potential. The Hamiltonian is
always of the {\it self-localizing type}, i.e. with a negative
nonlinear potential. We have study very recently two models with
this property in the framework of the ET problem\cite{yo}. The
simplest possible discrete symmetric donor-acceptor ET system,
including a nonlinear coupling of the Davydov type, is the
two-sites nonlinear Hubbard (NLH) model, given in second
quantization by the Hamiltonian:

\begin{eqnarray}
\hat{H}&=&\epsilon_{1} c_1^{\dagger}c_1 + \epsilon_{2}
c_2^{\dagger}c_2 +
t (c_2^{\dagger}c_1+ c_1^{\dagger}c_2)- \nonumber \\
& &\left[U \left<n_1\right>  c_1^{\dagger}c_1+ U \left<n_2\right>
c_2^{\dagger}c_2\right], \label{eqn11}
\end{eqnarray}
where $\left<n_i\right>=\left<c_i^{\dagger}c_i\right>$, and 1 (2)
refer to the donor (acceptor). The Hubbard-like terms $\propto
U>0$ give the necessary (non-linear) self-trapping potential and
$\epsilon_i$ and $t$ are the site energies and overlap (hoping)
integral between the two sites, respectively. This would be the
case of tunnelling between two equivalent sites with strong
polaronic coupling. This model, and variants of it including
dissipation, has been extensively studied for many years by Kenkre
and coworkers\cite{K1,K2,K3,K4} who coined the term {\it quantum
nonlinear dimer}. The objective of this Letter is to focus on a
specific property of the tunnelling between the two sites in this
model, namely: the presence of hysteresis. We shall then speculate
on situations where this property may be playing a role in ET with
nonlinear couplings. We review very briefly a few of its
properties discussed in Ref. \onlinecite{yo}. The ground state can
be sought by a Monte Carlo (MC) simulated annealing in $[z_1 -
z_2]$ space, were $z_1$ and $z_2$ are the complex coefficients of
the ground state wavefunction for the amplitude of the electron at
the donor and acceptor sites, respectively, i.e.

\begin{equation}
\Psi_G=\left(\matrix{z_{1}\cr
                     z_{2}}\right),
\label{eqn12}
\end{equation}

with $\left|z_1\right|^2+\left|z_2\right|^2=1$. A good measure of
the asymmetry in the electronic density of the two sites is the
magnitude

\begin{equation}
S\equiv \left|\left| z_1 \right|^2 - \left| z_2 \right|^2\right|,
\label{eqn13}
\end{equation}
which gives the absolute value of the difference between the
probability densities for donor and acceptor. Figure \ref{fig1} is
the calculated $S$ as a function of $t/U$ in the ground state of a
symmetric donor-acceptor system with $\epsilon_1=\epsilon_2=0$.
Alternatively, the minimization of (\ref{eqn11}) subject to the
normalization condition renders $S=\sqrt{1-(t/U)^2}$ for $t/U<1$
and $S=0$ for $t/U>1$, as seen in Fig. \ref{fig1}. If $t/U$ is
large (small $U$) the ground state is symmetric, i.e. $S=0$, and
the electron is equally shared between both sites. Below a certain
value of $t/U$, the system displays a {\it spontaneous symmetry
breaking} in the ground state, as discussed in Ref.
\onlinecite{yo}. From there on the system profits from
accumulating the wavefunction in one of the sites and
$S\rightarrow 1$ as $t/U\rightarrow 0$. This is the localization
in the symmetric donor-acceptor system mentioned by Knapp et.
al.\cite{Knapp}.

The question we raise now is how the electron behaves when we
start varying the energy $\epsilon_i$ of the levels externally. We
shall show that the intrinsic nonlinearity of Davydov-type models
for ET introduces hysteresis in the quantum mechanical tunnelling.

\begin{figure}[h]
 \centering{
  \includegraphics[width=8cm, height = 9cm]{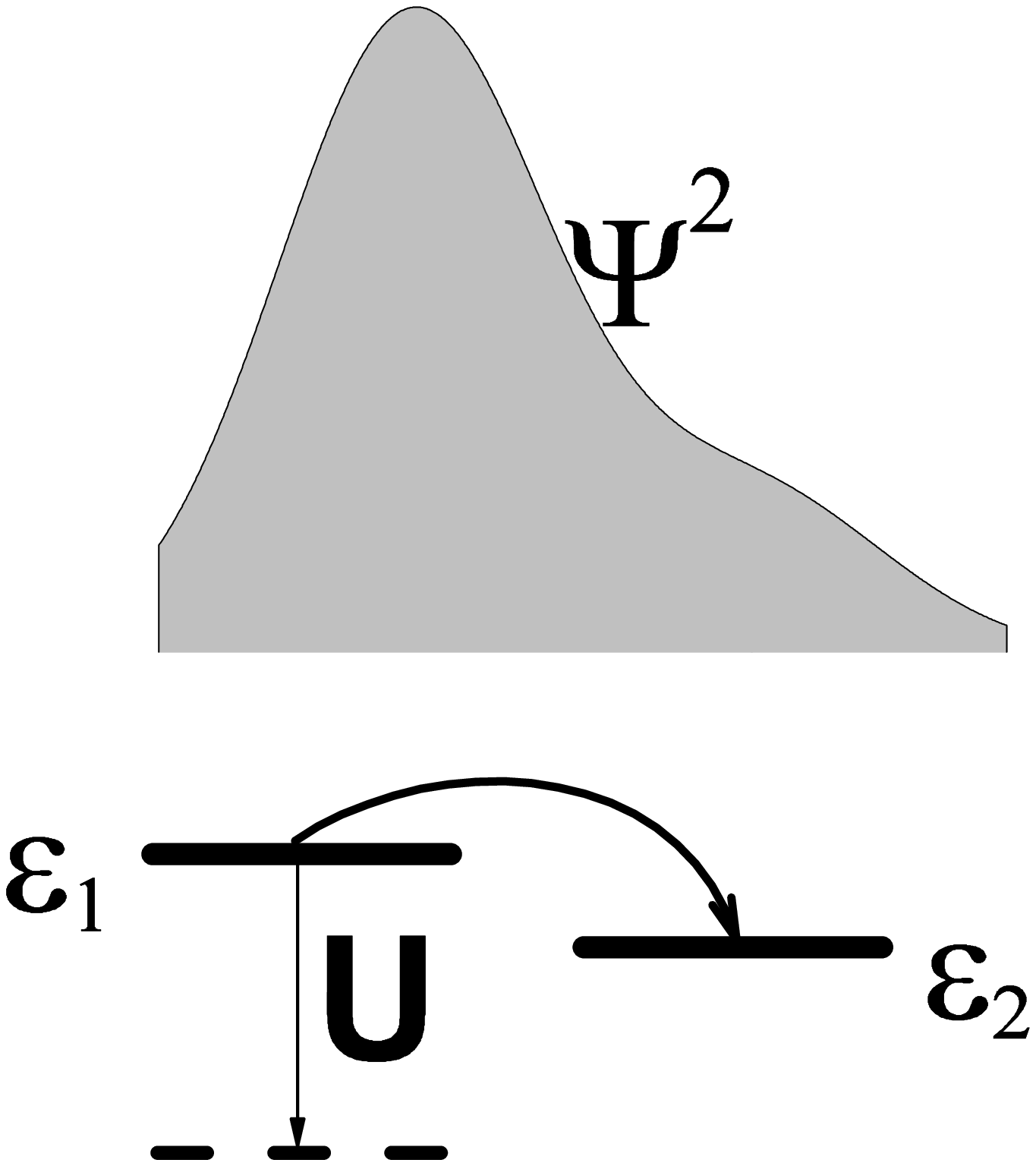}
 }
\caption{By assuming that the electron is initially localized on
the left (as shown schematically by the plot of $\psi^2$ on the
top), tunnelling to the right is hindered even if $\epsilon_2 <
\epsilon_1$ because of the correlation energy $U$. The system has
a memory of its ground state producing a hysteresis loop for the
tunnelling rate between both sites. See the text for further
details.}
 \label{fig2}
\end{figure}

The qualitative reason is relatively easy to understand and is
shown in Fig. \ref{fig2}. Let us assume that $U$ is big $(t/U$
small) and the electron is mainly {\it trapped} on the left. We
now vary the potential of $\epsilon_2$ making it lower than
$\epsilon_1$. Under normal circumstances, the electron would
tunnel immediately to the right. However this might not be the
case in the presence of a nonlinearity, because the energy of the
electron localized on the left is of the order of $\sim
\epsilon_1-U$, and this could be less than $\epsilon_2$, i.e. the
nonlinearity is holding the electron in its original place. There
is a memory effect, accordingly, which cannot exist in any linear,
adiabatic, quantum mechanical system. The same holds for the
reverse situation if we raise $\epsilon_2$ above $\epsilon_1$ once
the electron is localized on the right. By fixing $\epsilon_1$,
the tunnelling events left$\rightarrow$right or
right$\rightarrow$left occur at two different values of
$\epsilon_2$, producing an intrinsic hysteresis in a very simple
QM system.

 Mathematically, however, the property of hysteresis with a nonlinear
 Hamiltonian is in general difficult to prove. One could argue that in Fig. \ref{fig2},
 a gradual spread of the wavefunction from left to right is produced until the electron is
 fully localized on the right. And the opposite might hold on the
 reversed bias situation, thus producing a crossover without
 hysteresis. The way to prove the existence of hysteresis with a sudden
 transition between the two situations is by properly solving the
 model. Notwithstanding, a perturbative approach is doomed to fail.
 The reason is that a localized transition from donor to acceptor is more of a cooperative
 effect. In what follows we concentrate on the problem of the
 stability of the wavefunction given an initial (localized)
 starting condition and a bias between $\epsilon_1$ and
 $\epsilon_2$.

 It is advantageous to change to a continuous version of the model in Fig. \ref{fig2}. We treat a model with a similar
 property (spontaneous symmetry breaking) that is related to (\ref{eqn11}): the nonlinear particle
 in a box. In Ref. \onlinecite{yo} it was shown that a quantum
 mechanical particle in a box obeying the NLS equation undergoes
 an spontaneous symmetry breaking in the probability density for a
 sufficiently large value of the nonlinear potential. The particle
 is confined either on the left or the right of the box and it is
 possible to produce tunnelling between the two situations under
 the presence of an external stochastic potential. The NLS, which
 is the equivalent to (\ref{eqn11}) for a continuous system,
 reads

\begin{equation}
\gamma\nabla^2\psi(\underline{\bf r})+\beta
\left|\psi(\underline{\bf r})\right|^2\psi(\underline{\bf r})=E
\psi(\underline{\bf r}), \label{eqn5}
\end{equation}
where $\psi(\underline{\bf r})$ is subject to the boundary
conditions $\psi(\underline{\bf r})|_{\pm a/2}=0$, with $a=$ size
of the box. $E$ is the energy and $\gamma < 0$, $\beta< 0$ are the
{\it kinetic energy} and {\it self-trapping potential} parameters,
respectively. The relevant magnitude is the ratio
$\rho=|\gamma/\beta|$, fixing the relative strength of both. For
small $\rho$, an spontaneous self-localization of the wavefunction
either to the left or the right of the box occurs\cite{yo}, both
sides playing the equivalent roles of the donor-acceptor sites in
(\ref{eqn11}). The continuous model (\ref{eqn5}) allows the
inclusion of an electric field competing with the stability of the
self-localized ground state. The presence of an external
potential, in the one-dimensional problem, is added to
(\ref{eqn5}) through an additional term on the right-hand side of
the form:

\begin{equation}
V_0 \left(1-\frac{2x}{a}\right) \psi(x), \label{eqn66}
\end{equation}
where $V_0$ is a constant, and $\underline{\bf r}\equiv x$ in this
case. This potential produces a linear slope (constant electric
field) form one end $(x=-a/2)$ to the other $(x=+a/2)$ of the box.

In addition, the stability of the wavefunction can be tested with
the aid of the Fourier Monte-Carlo (FMC) method developed in Ref.
\onlinecite{yo}. We review here very briefly the basic concepts
involved, while leaving details to the references\cite{yo}. The
wavefunction can {\it always} be expressed by a Fourier series of
arbitrary length of the form

\begin{equation}
\psi(\underline{\bf
r})=\sqrt{\frac{(2/a)}{\sum\limits_{n}(a_{n}^2+b_{n}^2)}}
\sum\limits_{n}\left[a_n\cos(\underline{{\bf
k}}^c_n\cdot\underline{{\bf r}})+ b_n\sin(\underline{{\bf
k}}^s_n\cdot\underline{{\bf r}})\right], \label{eqn10}
\end{equation}
with $\underline{{\bf k}}^s_n =2\pi/a~(n+1)$ and $\underline{{\bf
k}}^c_n=2\pi/a~(n+1/2)$. $a_n$-$b_n$ are Fourier coefficients and
$\psi(\underline{\bf r})$ satisfies automatically both the
boundary and normalization conditions. The basic idea of the FMC
method is that the ground state is found through a simulated
annealing (as a function of an artificial temperature $T$) in
Fourier coefficient space $a_n$-$b_n$.

Once the ground state is found for $T=0$, the potential can be
changed and the evolution of the ground state monitored for small
finite temperature $T_0$. It is understood that the particle
follows adiabatically the changes in potential. The role of $T_0$
is that of a stochastic perturbation testing the stability of the
ground state; $T_0$ has to be smaller than the energy of any
metastability of the system in the presence of nonlinearities. The
convergence of the method is tested by doubling the number of
coefficients and comparing the quantitative results\cite{yo}.

\begin{figure}[h]
 \centering{
  \includegraphics[width=8cm, height = 12cm]{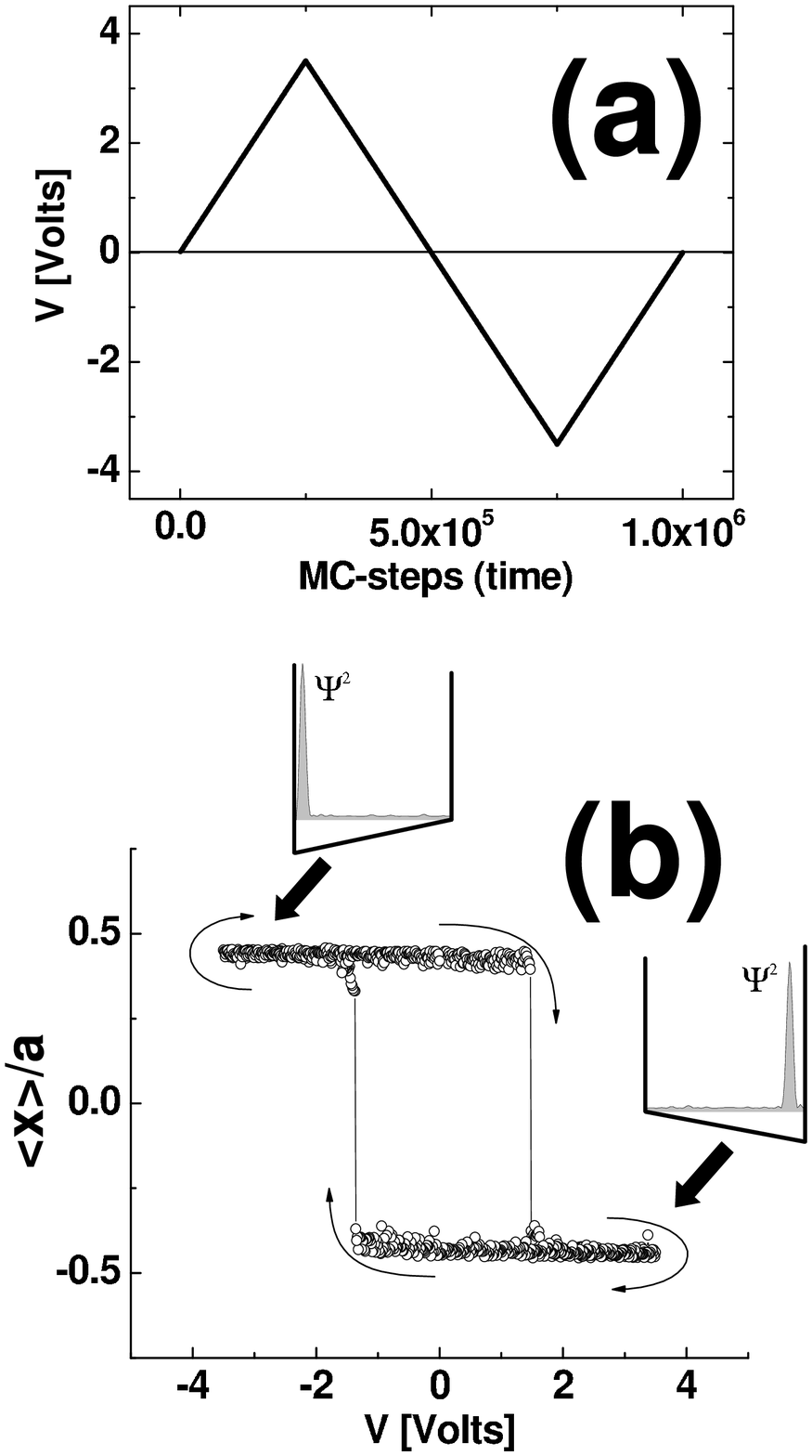}
 }
\caption{Hysteresis loop for the particle in a box with a
nonlinear potential (Eqn. \ref{eqn5}). In (a) the amplitude of the
potential is shown as a function of MC-steps, which play the role
of time. The parameters used in the simulation were $\gamma=-1.0$,
$\beta=-0.1$, and the size of the box $a=0.5$. Further details of
the numerical implementation of the FMC-method can be found in
Ref. \onlinecite{yo}. The particle follows adiabatically the
potential changes along the MC-changes. The initial state is with
the particle on the left, and it is found after a simulated
annealing with 20 temperatures from $T=1$ to 0, with 10$^4$
thermalization cycles per temperature and 40 Fourier coefficients
in total\cite{yo}. $T_0$ is fixed subsequently to 0.05 during the
potential changes. In (b) the evolution of the expectation value
of the wavepacket $\left<x\right>/a$ is plotted as a function of
the external potential. Note the presence of two very different
thresholds for the tunnelling in both directions. As in Fig.
\ref{fig2}, the reason is the competition between the intrinsic
nonlinearity of the Hamiltonian (once the particle is localized on
the left or right) and the applied external potential.}
 \label{fig3}
\end{figure}

Figure \ref{fig3} shows the result for a full cycle of the
electric field in both directions. The parameters of the
simulation are given in caption. We start with the particle
confined on the left (which we arbitrarily call donor) and we
follow the evolution of the wavepacket by monitoring
$\left<x\right>/a=(1/a)\int\limits_{-a/2}^{a/2} x \psi^2(x) {\rm
d}x$ for different potentials in the cycle shown in Fig.
\ref{fig3}(a). Figure \ref{fig3}(b) shows a result with the
hallmark of hysteresis. As before, the reason is the competition
between the intrinsic nonlinearity and the external potential,
fixing either the mismatch of the levels in Fig. \ref{fig2} or the
energy difference between left and right in Fig. \ref{fig3}(b).
The transition is not gradual (as it would be without
nonlinearity) because the electron does not profit from a
reduction in energy by transferring part of its wavefunction to
the other side, until a very specific condition in terms of the
competition with the correlation energy $U$ is achieved.

By having a model that combines basic aspect which are know to
exist in real ET systems, we can now go one step backwards and
contemplate those situations where this effect might be playing a
role. Control of ET is decisive in many biological functions; the
canonical example being ET in the mitochondrial chain, which must
stop when ADP is all phosphorylated. For more than 30 years it has
been suggested that the energy matching between donor and acceptor
could provide a mechanism of ET control\cite{DeVault2,Blumenfeld}.
In general terms, this ability of control has to be compatible
with the slowness of the reverse rate responsible for the near
100\% quantum efficiency observed, for example, in the
photosynthetic reaction centers, and with the presence of
couplings to the local environments. These requirements are
normally met by the so-called {\it energy-gap law} which
postulates two different activation barriers for the forward and
backward ET reactions, in the spirit of Marcus
theory\cite{Hall,DeVault,Marcus}. It has been also known for a
long time that simplified {\it in-vitro} model
systems\cite{Blankenship} invariably show similar or larger rates
for the backwards recombination. It seems then that vibronic
nonlinear couplings to the environment are {\it essential} to
understand the efficiency of the ET mechanism, and that evolution
through natural selection has engineered a favorable combination
of nonlinearity and control capabilities.

It is clear, accordingly, that a basic model like the one
presented here satisfies several of these conditions
simultaneously. The presence of QM-hysteresis is a key issue that
has been overlooked in the opinion of the present author, and
provides a very fine mechanism of control {\it even in the
presence of broadening of the electronic states through vibronic
coupling at the donor and acceptor sites}. In addition, the
back-reaction problem is automatically solved by the same token,
without having to resort to ideas which are normally more
difficult to justify, like the presence of two radically different
matrix elements for the forward and backward reaction,
respectively\cite{Jortner}. We believe that a model of tunnelling
between donor and acceptor in the spirit of the models proposed by
Davydov\cite{Davydov} and Knapp\cite{Knapp}, contain the essential
qualitative ingredients to explain simultaneously both control and
efficiency of real ET system found in biological systems. The main
objective of the present paper is then achieved in Fig.
\ref{fig3}, i.e. to demonstrate the existence of QM-hysteresis in
a very simple model and to highlight its possible relevance in a
variety of phenomena associated with ET in biological systems.

\begin{figure}[h]
 \centering{
  \includegraphics[width=8cm, height = 12cm]{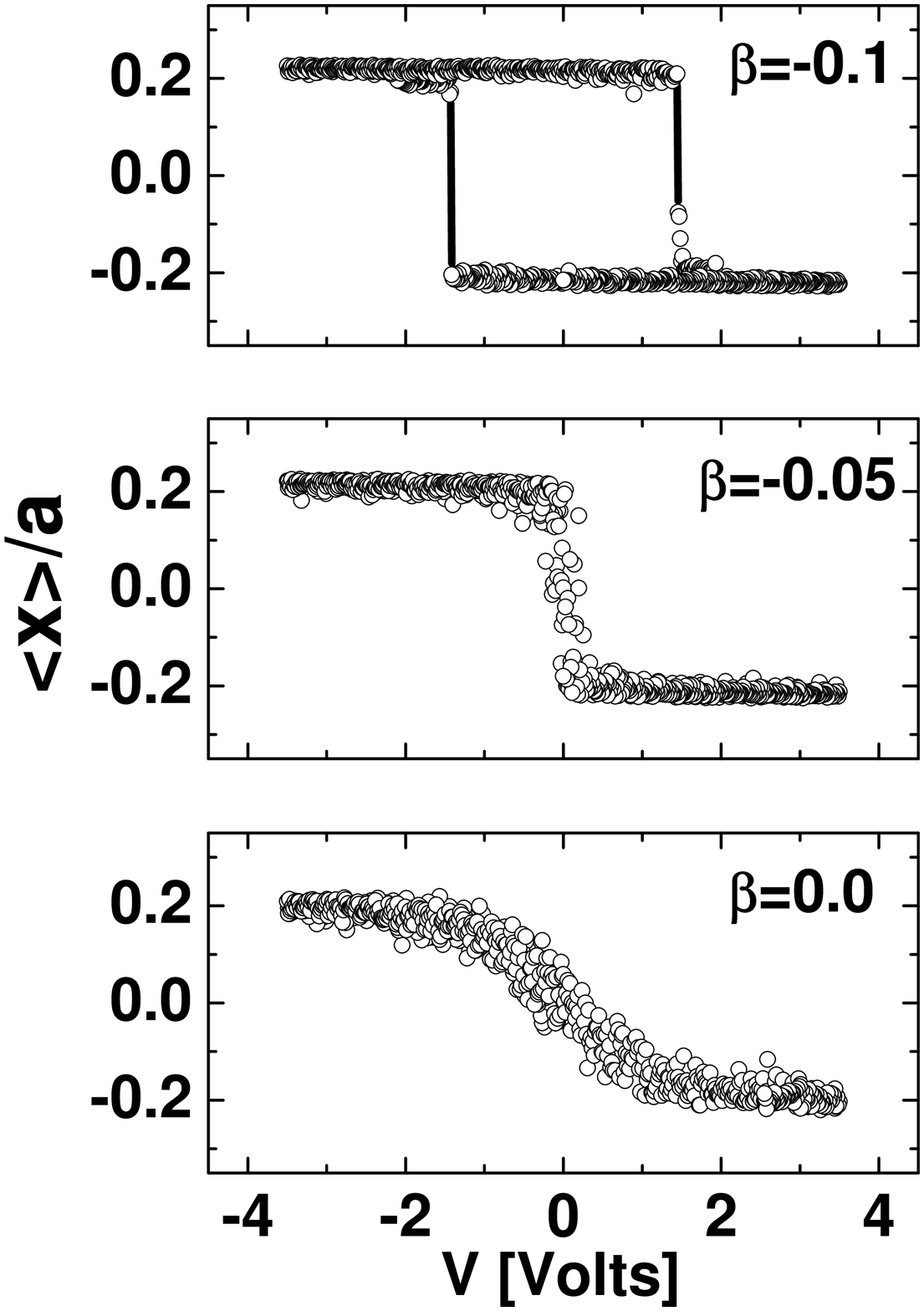}
 }
\caption{Hysteresis loops for different values of the nonlinear
coefficient $\beta$ in Eq. \ref{eqn5}. For a fixed $\gamma=-1.0$,
we show three examples of competition between nonlinearity and the
external electric field for $\beta=-0.1$ (top), $\beta=-0.0.5$
(middle), and $\beta=0.0$ (bottom). For $\beta=-0.05$ (middle) the
{\it coercive field} on the top graph has collapsed to a very
small value around $V\sim 0$. The system profits from accumulating
the probability density on the left or the right of the box, but
the barrier is very small and quickly overcome when the field
reverses. For $\beta=0.0$ the particle follows (linearly) the
electric field at low voltages, until the boundary conditions on
both sides of the box start competing (and saturating) the drive
of the external field. See the text for further details.}
 \label{fig4}
\end{figure}

The hysteresis cycle in Fig. \ref{fig3} defines a threshold value
for the transition between the two states, in analogy to the {\it
coercive field} displayed by magnetic systems. In order to show
that the hysteresis in the cycle shown in Fig. \ref{fig3} comes
effectively from a competition between the intrinsic nonlinearity
of the Hamiltonian and the external electric field, we show in
Fig. \ref{fig4} the effect of the parameters of the model on the
{\it coercive filed} of the transition. Figure \ref{fig4} shows
three examples for a fixed $\gamma$ and three different $\beta$'s:
-0.1, -0.05, and 0. For $\beta=-0.05$ the cycle has narrowed down
to a very small region around the origin. The system still profits
from accumulating the probability density on the left or the right
of the box, but jumps immediately form one situation to the other
under the slightest perturbation. The Fourier-MC method has
problems to define the exact width of the transition in this case,
because the perturbation parameter $T_{0}$ can be comparable to
the small barrier separating the two situations. For $\beta=0$,
the particle follows smoothly the external electric field. The
S-shaped average coordinate of the probability distribution is due
now to a competition between the boundary conditions on both sides
and the force exerted by the electric field. The wave-packet
follows the field linearly at very low fields, but as soon as the
displacement from the center is large enough, the boundary
conditions on the left or right of the box limit the response to
the field; thus achieving saturation in the displacement. These
results prove that the nonlinear bistable behavior in Fig.
\ref{fig3} is, effectively, a competition between nonlinearity and
external fields in the model.

On more speculative grounds, however, the existence of simple
QM-systems with tunnelling hysteresis at the molecular level
should also be of interest in other problems. For example, in the
field of molecular electronics (ME)\cite{Tour} where decisive
steps have been taken towards the production of logic devices
based on the change of electronic properties by single molecules.
Charge storage in a self-assembled molecular device has been
demonstrated already\cite{82b}; the device works by storing a high
or low conductivity state achieved by the addition or subtraction
of an electron. But the contact problem of organic molecules with
metals remains an outstanding problem in ME, and alternatives have
been proposed based on the so-called quantum cellular automata
(QCA) and electrostatic architectures\cite{Tour,144}. The QCA
approach has received considerable attention in ME as an
alternative to conventional electronic devices due to its
potential in reducing power dissipation by avoiding external
currents\cite{Tour}. It is also an interesting conceptual shift
from conventional electronic design. After all, the most
sophisticated machines known in nature, which are living organism,
work mainly not through changes in conductivity and currents but
rather through electrostatic and van der Waals interactions.

Even though the technology to achieve ME and QCA devices is only
beginning to appear, the results in this paper suggest that
organic symmetric donor-acceptor pairs with strong polaronic
coupling, and external control by an electric field, are excellent
candidates for the ultimate single-molecule dynamic random access
memory (RAM) device.

Partial support from EPRSC (UK) at Imperial College London (where
part of the work has been done) is gratefully acknowledged.

\newpage

\end{document}